# Beam halo study on ATF damping ring [#]

Dou Wang[1,*], Philip Bambade[2], Kaoru Yokoya[3], Takashi Naito[3], Jie Gao[1]
[1]IHEP, Beijing 100049, China
[2]LAL, Osay, 91898, France
[3]KEK, Tsukuba, 305-0801, Japan

Abstract

Halo distribution is a key topic for background study. This paper has developed an analytical method to give an estimation of ATF beam halo distribution. The equilibrium particle distribution of the beam tail in the ATF damping ring is calculated analytically with different emittance and different vacuum degree. The analytical results agree the measurements very well. This is a general method which can be applied to any electron rings.

## I. Introduction

The distribution function of an electron bunch, transverse or longitudinal, is often assumed to be Gaussian. Actually, however, due to stochastic processes, there always exists some deviation and hence charge distributions of accelerator beams can be separated into two parts: the beam cores, which usually have Gaussian-like distributions, and the beam halos, which have much broader distributions than the beam cores. The central part affects the luminosity of colliders, circular or linear, and the brightness of synchrotron light sources, while the halos can give rise to background in collision experiment detectors and even reduce the lifetime if its distribution is too large.

Both calculations and measurements for beam halo are a hard topic. For the RMS emittance growth, we do have some mature theories and numerical codes to use, while for the whole beam distribution, especially for the halo part, there are few mature theories. Even using simulations, it's still difficult to get the halo distribution with three dimensions because the beam halo includes much fewer particles than the beam core. For the first time, we have developed a series of theory to estimate the whole beam distribution, including the halo section based on the theory established by K. Hirata and K. Yokoya [1]. We have focused on three main mechanisms: beam-gas scattering, beam-gas bremsstrahlung and intrabeam scattering. Beam-gas scattering produces transverse halo, beam-gas bremsstrahlung produces longitudinal halo, and intrabeam scattering can induce both transverse and longitudinal halo.

At the interaction point (IP) of ATF2 (Accelerator Test Facility 2), an elaborately designed beam size monitor based on laser interferometer technology, called the Shintake monitor, is utilized to measure the sub-100 nm electron beam size [2]. However, the photon background in the IP section will influence the modulation of the Shintake monitor, and hence degrade the resolution of beam size measurements. So the beam halo distribution is important for the measurement of the beam size at IP. Since Understanding charge distribution of the beam halo and how it is created are essential to our estimation of background. We have made some analytical estimation of halo distribution in ATF due to three common stochastic processes using our own theories. (Typical ATF damping ring parameters are listed in Table 1.) Also, we have compared the theoretical estimation with the newest measurements by advanced halo monitor. The analytical results agree the measurements very well.

___________
[#] Work supported by National Key Programme for S&T Research and Development (Grant NO.: 2016YFA0400400) and the National Foundation of Natural Sciences (11505198 and 11575218).
[*] Email: wangd93@ihep.ac.cn

Table 1: Typical ATF parameters

| Parameter | Value |
|---|---|
| Energy $E_0$ (GeV) | 1.3 |
| Natural energy spread $\delta_0$ | $5.44\times10^{-4}$ |
| Energy acceptance | 0.005 |
| Average $\beta x/\beta y$ (m) | 3.9/4.5 |
| Horizontal emittance (nm) | 1.3 |
| Vertical emittance (pm) | 20 |
| Transverse damping time (ms) | 18.2/29.2 |
| Longitudinal damping time (ms) | 20.9 |

## II. Theory review
### A. Beam-gas scattering

The performance of accelerators and storage rings depends on the many components of the accelerator, and one very important component is the vacuum system. Interactions between the accelerated particles and the residual gas atoms may degrade the beam quality. The lifetime may be reduced and/or the emittance may increase. The beam halo is possibly generated because the particles' distribution deviates from a Gaussian distribution.

The deflection of an electron via the Coulomb interaction is described by Rutherford scattering. We assume that this scattering is elastic and that the recoil momentum of the residual gas is negligible. The differential cross-section of the electron scattering with an atom is given by [3]

$$\frac{d\sigma}{d\Omega} = \left(\frac{2Zr_e}{\gamma}\right)^2 \frac{1}{(\theta^2 + \theta_{\min}^2)^2} \tag{1}$$

where Z is the atomic number, $r_e$ is the classical electron radius, $\gamma$ is the relativistic Lorentz factor and $\theta_{\min}$ is the minimum scattering angle which is determined by the uncertainty principle as

$$\theta_{\min} = \frac{Z^{1/3}\alpha}{\gamma} \tag{2}$$

where $\alpha$ is the fine structure constant. If we integrate over the whole space angle $\Omega$, we obtain the total cross-section

$$\sigma_{tot} = \int_0^{2\pi}\int_{\theta_{\min}}^{\pi} \left(\frac{2Zr_e}{\gamma}\right)^2 \frac{1}{\left(\theta^2 + \theta_{\min}^2\right)^2} \sin\theta\, d\theta\, d\varphi$$
$$\approx 4\pi Z^{4/3}(192 r_e)^2 \tag{3}$$

We then need to get the probability density function $f(\theta)$. Assuming $\theta^2 = \theta_x^2 + \theta_y^2$, then integrating over one direction will give the differential cross-section for the other direction

$$\frac{d\sigma}{d\theta} = \frac{4\pi r_e^2 Z^2}{\gamma^2} \frac{1}{\left(\theta^2 + \theta_{\min}^2\right)^{3/2}} \tag{4}$$

Here and hereafter we denote

$$\theta \equiv \theta_x(\theta_y) \tag{5}$$

Thus,

$$f(\theta) = \frac{1}{\sigma_{tot}} \frac{d\sigma}{d\theta} = \frac{\theta_{min}^2}{\left(\theta^2 + \theta_{min}^2\right)^{3/2}} \tag{6}$$

$$\left(\int_0^\infty f(\theta)d\theta = 1\right)$$

For the elastic scattering, we assume that CO gas is dominant for beam-gas scattering, so that the total scattering probability in a unit time is

$$N = Q\sigma_{tot} c \tag{7}$$

where $c$ is the speed of light and $Q$ is the number of gas molecules in a unit volume, given by

$$Q = 2.65 \times 10^{20} nP \tag{8}$$

where $n$ is the number of atoms in each gas molecule and $P$ is the partial pressure of the gas in pascals. (Here for CO gas, $Z=50^{1/2}$ and $n=2$).

The collision probability of electron and gas atoms during one damping time is

$$N_\tau = N\tau \tag{9}$$

where $\tau$ is the transverse damping time for either the horizontal or vertical direction.

Finally, one gets the beam transverse distribution as

$$\begin{aligned}\rho(X) &= \frac{1}{\pi}\int_0^\infty \cos(kX)\exp[-\frac{k^2}{2} + \frac{2}{\pi}N_\tau \times \int_0^1 \frac{\left(\int_0^\infty \cos(\frac{k}{\sigma_0'}x\theta)f(\theta)d\theta\right) - 1}{x}\arccos(x)dx]dk \\ &= \frac{1}{\pi}\int_0^\infty \cos(kX)\exp[-\frac{k^2}{2} + \frac{2}{\pi}N_\tau \times \int_0^1 \frac{\Theta xk K_1(\Theta xk) - 1}{x}\arccos(x)dx]dk\end{aligned} \tag{10}$$

where $\Theta$ is the minimum scattering angle normalized by angular beam size, which is defined by $\frac{\theta_{min}}{\sigma_0'}\left(\sigma_0' = \frac{\sigma_0}{\beta}\right)$, and $X$ denotes both horizontal and vertical normalized coordinate. This formula tells us that the beam distribution disturbed by the beam-gas scattering effect is decided by only two parameters, the normalized scattering frequency $N_\tau$ and the normalized minimum scattering angle $\Theta$.

**B. Beam-gas bremsstrahlung**

As is well known, when charged particles are accelerated, they emit electromagnetic radiation, i.e. photons. In accelerators, an electron with energy $E_0$, which passes a molecule of the residual gas, is deflected in the electric field of the atomic nucleus. The electron loses energy due to the radiation emitted when an electron is deflected. This bremsstrahlung will be very strong for relativistic electrons. There is a certain probability that a photon with energy $\varepsilon$ is emitted and the differential cross-section for an energy loss $\varepsilon$ due to bremsstrahlung is given by [4]

$$\frac{d\sigma}{d\varepsilon} = 4\alpha r_e^2 Z(Z+1)(\frac{4}{3}\ln\frac{183}{Z^{1/3}} + \frac{1}{9})\frac{1}{\varepsilon} \tag{11}$$

Then, one can get the total scattering frequency

$$\begin{aligned}\sigma_{tot} &= \int_{E_{min}}^{E_{max}} \frac{d\sigma}{d\varepsilon}d\varepsilon = 4\alpha r_e^2 Z(Z+1) \\ &\times (\frac{4}{3}\ln\frac{183}{Z^{1/3}} + \frac{1}{9})\ln\frac{E_{max}}{E_{min}}\end{aligned} \tag{12}$$

and the probability density function

$$f(\varepsilon) = \frac{1}{\sigma_{tot}} \frac{d\sigma}{d\varepsilon} = \frac{1}{\ln\frac{E_{max}}{E_{min}}} \frac{1}{\varepsilon} \qquad \left( \int_{E_{min}}^{E_{max}} f(\varepsilon)d\varepsilon = \right) \tag{13}$$

where $E_{max}$ is the maximum energy loss which equal to the ring energy acceptance and $E_{min}$ is the minimum energy loss for each scattering.

Also, using the same formulae given in Eq. (7) to Eq. (9), we can calculate the total collision frequency.

Thus, the beam energy distribution due to beam-gas bremstruhlung can be expressed by

$$\rho(E) = \frac{1}{\pi} \int_0^\infty \cos(kE) \exp[-\frac{k^2}{2} + \frac{2}{\pi} N_\tau \times \int_0^1 \frac{\left( \int_{E_{min}}^{E_{max}} \frac{\cos(\frac{kx}{E_0\delta_0}\varepsilon)}{\left(\ln\frac{E_{max}}{E_{min}}\right)\varepsilon} d\varepsilon \right) - 1}{x} \arccos(x)dx]dk \tag{14}$$

### C. Intra-beam scattering

Intra-beam scattering (IBS) is the result of multiple small-angle Coulomb collisions between particles in the beam, which is different from the Touschek effect. The Touschek effect describes collision processes which lead to the loss of both colliding particles. In reality, however, there are many other collisions with only small exchanges of momentum. Due to the scattering effect, beam particles can transform their transverse momenta into longitudinal momenta randomly, which leads to a continuous increase of beam dimensions and to a reduction of the beam lifetime when the particles hit the aperture. Detailed theories of intra-beam scattering have been developed in references [5-11]. However, the existing theories mainly discuss the rms emittance growth and the rise time due to intra-beam scattering, which cannot give the whole information of particle distribution because the real beam has a non-Gaussian distribution. In this paper, we will discuss the IBS induced beam dilution for the longitudinal and vertical directions.

In the center-of-mass reference frame of two scattering particles, the differential cross section of Coulomb scattering for electrons (or positrons) is given by the Möller formula [12]

$$\frac{d\bar{\sigma}}{d\Omega} = \frac{4r_e^2}{(v/c)^4} \left( \frac{4}{\sin^4\theta} - \frac{3}{\sin^2\theta} \right) \tag{15}$$

where $v$ is the relative velocity in the c. m. system that we will assume to be essentially horizontal because horizontal momentum is much larger than vertical momentum and hence will contribute more to momentum exchange, and $\theta$ is the scattering angle. The bar denotes the center-of-mass reference frame and the differential cross section $d\bar{\sigma}$ is evaluated in the center-of-mass system.

At small angles (as is common for IBS), the Möller formula for the differential cross section reduces to

$$\frac{d\bar{\sigma}}{d\Omega} = \frac{16r_e^2}{(v/c)^4} \frac{1}{\sin^4\theta} \tag{16}$$

Considering the angular change of the momentum gives a momentum component perpendicular to the horizontal axis

$$p_\perp = p_x \sin\theta \tag{17}$$

and

$$dp_\perp = p_x \cos\theta d\theta \approx -p_x d\theta \tag{18}$$

with

$$p_x = \frac{m_0 v}{2} \tag{19}$$

where $m_0$ is the rest mass of the electron. Also considering

$$d\Omega = 2\pi \sin\theta d\theta \tag{20}$$

one can get

$$d\bar{\sigma} = 2\pi \frac{r_e^2}{\bar{\beta}^2} \frac{dp_\perp}{p_\perp^3} \tag{21}$$

where $\bar{\beta}$ is the c.m. velocity of the electrons in units of $c$ ($\bar{\beta} = \frac{v}{2c}$) and $p_\perp$ is the momentum exchange from the horizontal direction to the perpendicular directions in the center-of-mass frame.

Furthermore, taking account of the fact that the probability is the same for transfers occurring in the vertical and longitudinal directions, we can get the differential cross section for longitudinal momentum growth in the center-of-mass system

$$d\bar{\sigma} = \pi \frac{r_e^2}{\bar{\beta}^2} \frac{d\varepsilon}{\varepsilon^3} \tag{22}$$

where $\varepsilon$ is the longitudinal momentum change due to the IBS effect in the center-of-mass system. If we transfer the longitudinal momenta back to the laboratory system, the real longitudinal momentum growth will be $\gamma\varepsilon$.

Finally, for a single test particle, the total number of events of momentum exchange from the horizontal direction to the longitudinal direction per second and the probability density function $f(\varepsilon)$ can be written as Eq. (23) and Eq. (24).

$$N = \frac{4\pi}{\gamma^2} \int \bar{\beta} c P(\vec{x}_1, \vec{x}_2) \int_{E_{\min}}^{\infty} \frac{d\bar{\sigma}}{d\varepsilon} d\varepsilon d\vec{x}_1 d\vec{x}_2 \approx \frac{c r_e^2}{6\gamma^3} \frac{N_e}{E_{\min}^2 \sigma_x \sigma_y \sigma_z \sigma_{x'}} \tag{23}$$

$$f(\varepsilon) = \frac{1}{\bar{\sigma}_{tot}} \frac{d\bar{\sigma}}{d\varepsilon} = \frac{2E_{\min}^2}{\varepsilon^3} \tag{24}$$

For the integration of Eq. (23), we have used the approximate result in reference [9].

Thus, one gets the expression of beam energy distribution due to the IBS process as

$$\rho(E) = \frac{1}{\pi} \int_0^\infty \cos(kE) \exp[-\frac{k^2}{2} + \frac{2}{\pi} N_\tau \times \int_0^1 \frac{(\int_{E_{\min}}^\infty \frac{2E_{\min}^2 \cos(\frac{kx}{E_0 \delta_0} \gamma\varepsilon)}{\varepsilon^3} d\varepsilon) - 1}{x} \arccos(x) dx] dk \tag{25}$$

where $N_\tau$ is the total scattering rate normalized by the longitudinal damping rate ($N_\tau = N\tau_z$) and $E_{\min}$ is the minimum momentum increment in the longitudinal direction during IBS process.

Furthermore, using the same method of beam-gas scattering, one can get the vertical distribution due to IBS as

$$\rho(Y) = \frac{1}{\pi}\int_0^\infty \cos(kY)\exp[-\frac{k^2}{2}+\frac{2}{\pi}N_\tau \times \int_0^1 \frac{(\int_{P_{min}}^\infty \frac{2P_{min}^2 \cos(\frac{kx}{\sigma_y}p_y)}{p_y^3}dp_y)-1}{x}\arccos(x)dx]dk \qquad (26)$$

where $N_\tau$ is the total scattering rate normalized by the vertical damping rate ($N_\tau = N\tau_y$) and $P_{min}$ is the minimum momentum increment in the vertical direction during the IBS process.

### III. Analytical estimation for ATF beam halo
#### A. Beam-gas scattering

According to Eq. (10), we calculated the beam halo distribution with different emittance and different vacuum degree.

- halo distribution with different vacuum pressures ($E_0$=1.3GeV, $\varepsilon_x$=1.3nm, $\varepsilon_y$=20pm)

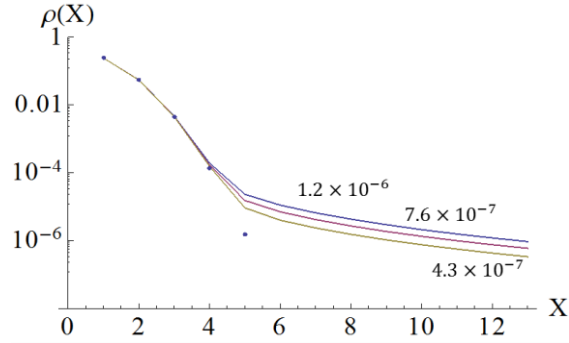

Figure 1: Horizontal beam distribution with different vacuum pressures (horizontal coordinate $X$ is normalized by RMS beam size).

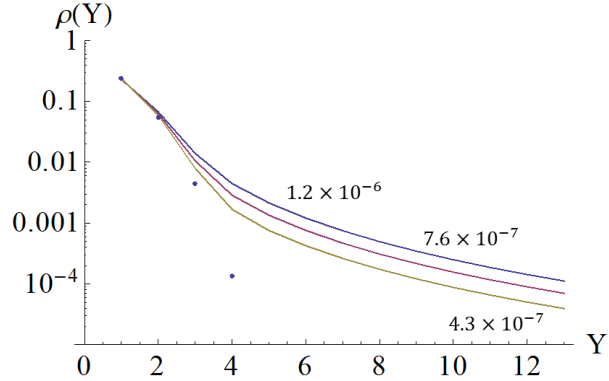

Figure 2: Vertical beam distribution with different vacuum pressures (vertical coordinate $Y$ is normalized by RMS beam size).

From Fig. 1 and Fig. 2, we can see that due to the beam-gas scattering effect, the beam distribution will deviate from a Gaussian distribution. Worse vacuum status will give a larger beam halo and smaller Gaussian beam core. Also, it can be seen that the vertical distribution of a beam is affected more than the horizontal distribution by the elastic beam-gas scattering because $\sigma_{y0}'<<\sigma_{x0}'$, so $\Theta_y>>\Theta_x$.

- halo distribution with different emittance ($E_0$=1.3GeV, $P$= $10^{-6}$ Pa)

Fig. 3 and Fig. 4 show that larger emittance will give smaller halo.

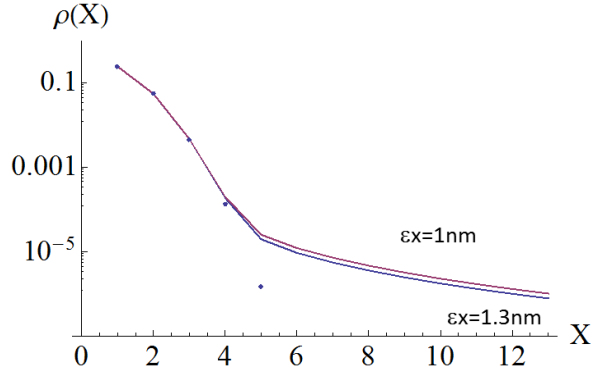

Figure 3: Horizontal beam distribution with different emittance (horizontal coordinate X is normalized by RMS beam size).

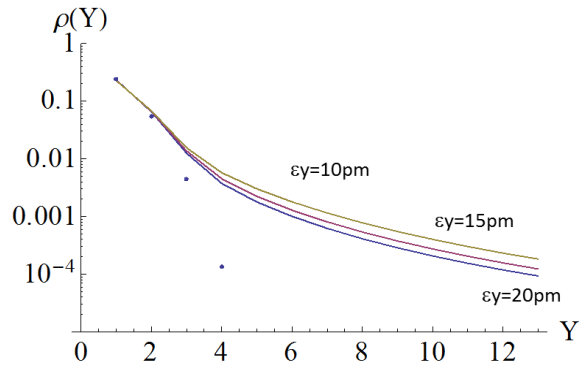

Figure 4: Vertical beam distribution with different emittance (vertical coordinate Y is normalized by RMS beam size).

### B. Beam-gas bremsstrahlung

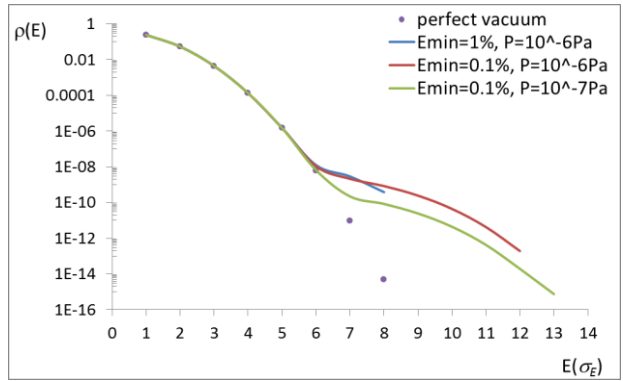

Figure 5: Energy distribution with different vacuum pressures and different minimum energy loss (The horizontal coordinate E is normalized by the natural energy spread).

Fig. 5 shows the beam energy distribution based on Eq. (14). It can be seen that the level of beam halo is decided by the purity of the vacuum. Lower vacuum pressure will give a smaller beam halo. Also, it shows that minimum energy loss per scattering $E_{min}$ is an important parameter which can be adjusted. We therefore need to choose an appropriate $E_{min}$, keeping in mind a balance of CPU computing time and halo length.

### C. Intra-beam scattering

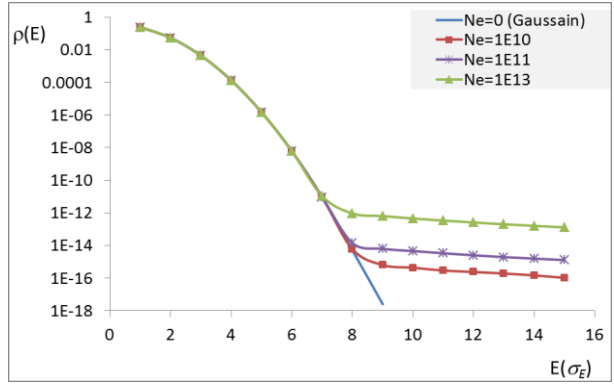

Figure 6: Energy distribution with different bunch populations (The horizontal coordinate $E$ is normalized by the natural energy spread).

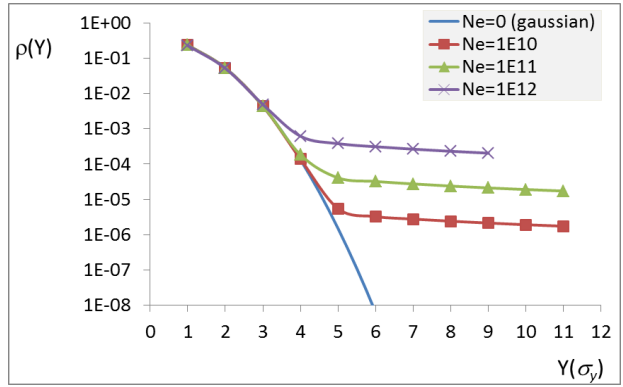

Figure 7: Vertical distribution with different bunch populations (vertical coordinate Y is normalized by RMS beam size).

Fig. 6 shows the beam energy distribution based on Eq. (25). Here, we choose $E_{min}$ equal to 0.01% of nature energy spread. We can see that a larger beam density give a larger beam halo, which will also increase the RMS beam size. Since the design bunch population is $1\times10^{10}$ for the ATF damping ring, from Fig. 6, the beam energy distribution will deviate from a Gaussian shape outside $8\sigma_E$ and the halo particles will have about $1\times10^{-16}$ of peak beam density. Compared with Fig. 5, it can be seen that in the ATF damping ring, the energy distribution of the beam halo is dominated by the beam-gas bremsstrahlung effect rather than the IBS effect.

Fig. 7 shows the vertical charge distribution based on Eq. (26). Here, we choose $P_{min}$ about 0.02% of the natural energy spread. In the ATF damping ring, the vacuum level is at the order of $10^{-7}$ ~$10^{-6}$ Pa. According to Fig. 2, the charge intensity of the vertical halo is about 4 orders of magnitude lower than the beam core in the ATF due to beam-gas scattering effect. So it seems that in the ATF damping ring, the vertical distribution is dominated by beam-gas scattering rather than by the IBS effect.

## IV. Comparison with measurements
### A. Measurement with advanced halo monitor

In order to measure the beam halo distribution and make comparison with analytical estimation, KEK-ATF2 developed a beam halo monitor which has both high resolution and high sensitivity based on fluorescence screen. A YAG: Ce screen, which has 1 mm slit in the center was set in the beam line. The image on fluorescence screen is observed by imaging lens system and CCD camera. In this configuration, the beam in the core will pass through the slit. The beam in surrounding halo will hit the fluorescence screen, and we can observe the distribution of beam halo. The intensity contrast of beam halo to the beam core is measured by scanning the beam position for the fixed fluorescence screen position. Fig. 8 and Fig. 9 show the very fresh measurements in 2015 by A YAG: Ce screen [13].

By comparing Fig. 8 and Fig. 9 with Fig. 2, we found amazing agreement. Also we can see little difference between Fig. 8 and Fig. 9. It is a good proof of our prediction that in ATF the vertical distribution is dominated by beam-gas scattering rather than by the IBS effect.

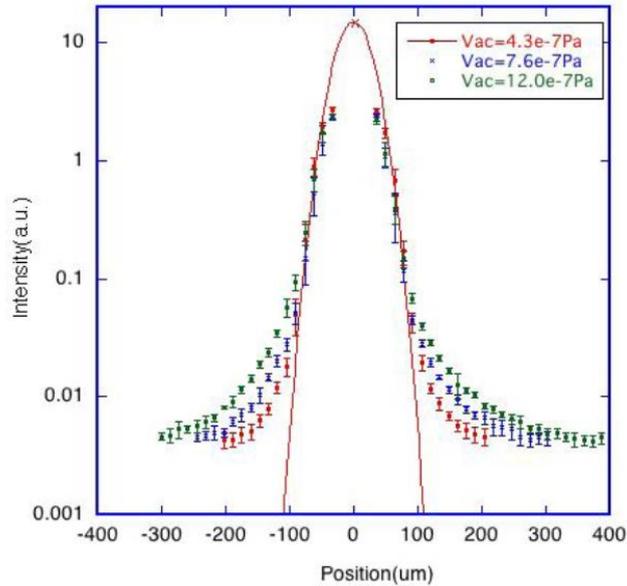

Figure 8: Vertical distribution of the beam halo for the different vacuum condition in the case of the beam intensity $0.23 \times 10^{10}$ electrons

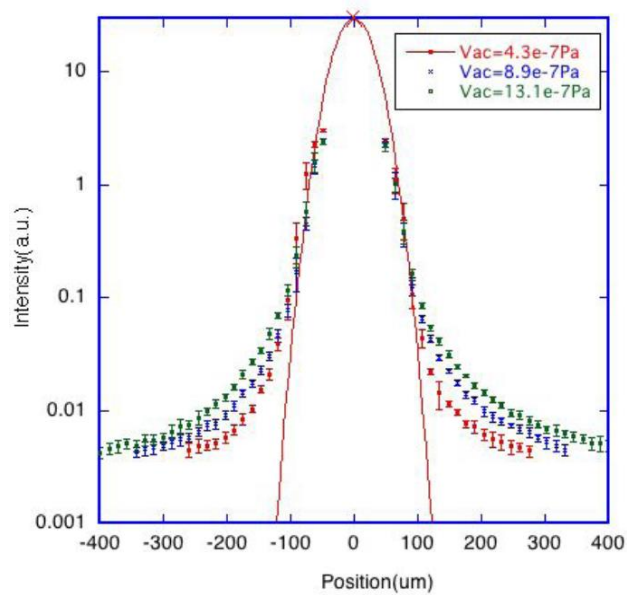

Figure 9: Vertical distribution of the beam halo for the different vacuum condition in the case of the beam intensity $0.45 \times 10^{10}$ electrons

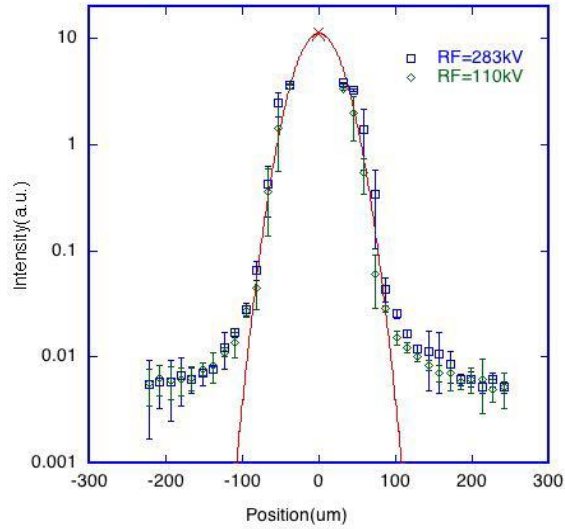

Figure 10: Distribution of ATF beam halo for the different RF voltage in the case of the beam intensity $0.23 \times 10^{10}$ electrons

The distribution of the beam halo for the different RF voltage of the damping ring is plotted in Figure 10 in the case of the beam intensity $0.23\times10^{10}$ electrons [13]. Two different RF voltages are plotted, Vrf=283kV and Vrf=110kV, respectively. The bunch length and the energy spread of the beam is a function of the RF voltage and the difference is about 10% for the two cases. The bunch length will affect the core beam size by the intra-beam scattering. The measurement shows almost same beam halo and a little bit increased the beam core. Once again it verified our theoretical expectation that ATF vertical halo is determined by beam-gas scattering effect rather than IBS effect.

### B. Measurement with wire scanner

Before the completion of ATF2 beam line, halo distribution was also measured in the old ATF extraction line with wire scanners in ATF spring run of 2005 [2]. The wire scanner consists of a metal wire (tungsten in the ATF) with a micromover to scatter the electron beam at every wire position. The scattered photons are counted by a gamma detector, which is an air-Cherenkov counter with a 2 mm thick lead converter and a PMT is attached for the photon counting.

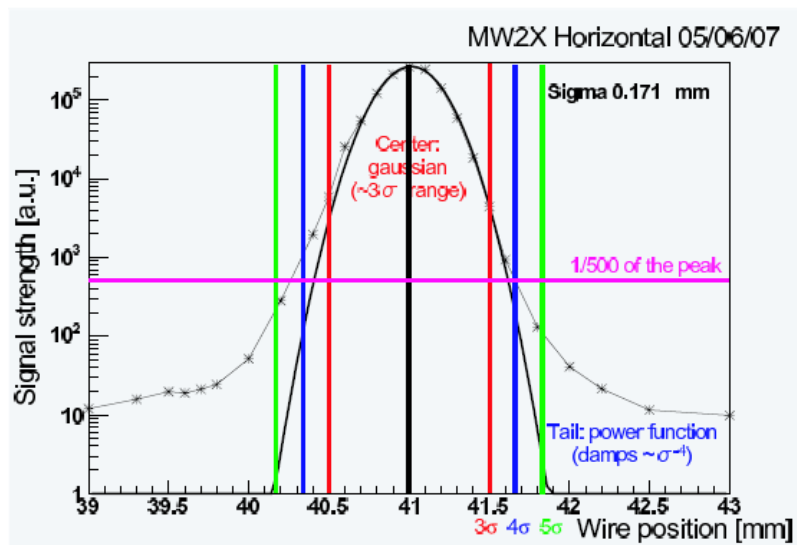

Figure 11: Horizontal charge distribution using the ATF extraction line wire scanner MW2X.

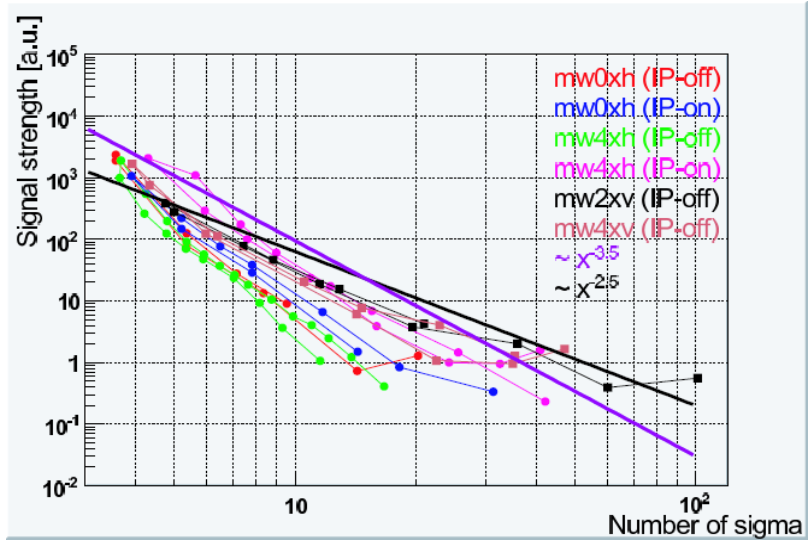

Figure 12: Measurement of the halo part using several wire scanners for both vertical and horizontal directions. (Horizontal axis is normalized by beam size.) Vertical beam profiles are shown as a square, and horizontal beam profiles are shown as a circle. The difference of the IP-on data and the IP-off data is the vacuum level. For the IP-off data, some of the ion pumps in the ATF dumping ring were turned off to obtain data with degraded vacuum. The difference of the vacuum level is about 1:5.

Fig. 11 is the horizontal charge distribution using the ATF extraction line wire scanner MW2X. The plot shows that the distribution in the beam center of $< 4\sigma$ range is well approximated by a Gaussian (bold line), while in the region of $> 4\sigma$, the deviation from the central Gaussian is large. This measurement result agrees well with analytical estimation in Fig. 1 and Fig. 3.

Fig. 12 is the halo distribution for both horizontal and vertical directions which are measured at different locations. This plot shows a comparison of the halo distribution for several beam sizes. It is also a proof to our theoretical expectation that the vertical distribution of the beam is affected more than the horizontal distribution due to beam-gas scattering

## V. Remaining issues

There are still several issues remaining to be addressed for a whole systematic study.

- The horizontal halo due to IBS, where a coupling effect between longitudinal direction and horizontal direction exists through horizontal dispersion, has not been solved out theoretically. The horizontal distribution due to IBS will be more difficult than the vertical direction.
- The combined influence on beam halo due to beam-gas scattering ring, beam-gas bremsstrahlung and IBS has not been given.
- It is a pity we have no measurement results for longitudinal distribution to support our theory related to the longitudinal halo so far.

## VI. conclusions

Due to various incoherent stochastic processes in the electron (positron) rings in an accelerator, the beam distribution will deviate from a Gaussian shape, generating a longer beam tail and increasing the beam dimensions. With the background issue, we have to study the halo distributions and the mechanisms by which the halo particles are produced. Once we understand the mechanisms of how the halo comes up, we can estimate the intensity level at halo part and we know how to control the halo even to reduce halo, also we can provide the vacuum requirement during machine design stage in order to control the beam halo at certain level. Take ATF as an example, we try to estimate the halo status with different emittance and vacuum level. In this paper, we have calculated the whole beam distribution of the ATF damping ring, including the halo section, based on our own theory. By comparing with measurements, we saw a good agreement between the analytical method and the experimental results. The analytical method developed in this paper is not specific to ATF and can be utilized on any circular electron (positron) accelerator.

For the RMS emittance growth, we do have some mature theories and numerical codes to use, while for the whole beam distribution, especially for the halo part, there are few mature theories. Even using simulations, it's still difficult to get the halo distribution for three directions because the beam halo includes much fewer particles than the beam core. For the first time, we have given a theoretical method to estimate the whole beam distribution of the lepton ring, including the halo section, with different emittance and vacuum level. From our study, we know that the transverse halo in ATF is dominated by beam gas scattering, and also smaller emittance and worse vacuum give larger beam halo. Also we can expect that the longitudinal halo in ATF is dominated by the beam-gas bremsstrahlung effect. For the next, we are trying to study the horizontal distribution due to IBS and the method how to evaluate the combine effect from the three stochastic processes.